\documentclass[12pt]{article}
\pdfoutput=1 
\usepackage{graphicx} 
\usepackage{color}      
\usepackage{cancel,tabularx,moreverb,fancybox,amsmath,float,bm,braket,slashbox,txfonts,amssymb,bm,accents,enumerate}
\usepackage[top=30truemm,bottom=30truemm,left=25truemm,right=25truemm]{geometry}
\usepackage{latexsym}
\usepackage{here}
\usepackage{cite}
\usepackage{hyperref}         

\newcommand{\ex}[1]{\mathrm{e}^{#1}}

\newcommand{\pa}[1]{\left(#1 \right)}

\newcommand{\ca}[1]{\mathcal{#1}}

\newcommand{\abs}[1]{\left|#1\right|}

\newcommand{\ti}[1]{\tilde{#1}}

\newcommand{\fr}{\frac}

\def\be{\begin{equation}}
\def\ee{\end{equation}}
\def\ba{\begin{eqnarray}}
\def\ea{\end{eqnarray}}

 \def\ba{{\bar{\alpha}}}

\def\tr{{\text{tr}}}

  \makeatletter
    
    \@addtoreset{equation}{section}
  \makeatother

\begin{document}

\begin{titlepage}
\thispagestyle{empty}

\begin{flushright}
CALT-TH 2023-031
\\
IPMU 23-0025
\\
RIKEN-iTHEMS-Report-23
\\

\end{flushright}

\bigskip

\begin{center}
\noindent{{\large \textbf{
Universality of Effective Central Charge in Interface CFTs
}}}\\
\vspace{2cm}
Andreas Karch,${}^1$
Yuya Kusuki,${}^{2,3}$
Hirosi Ooguri,${}^{2,4}$
Hao-Yu Sun,${}^1$
Mianqi Wang${}^1$
\vspace{1cm}

${}^1${\small \sl Theory Group, Weinberg Institute, Department of Physics \\ University of Texas, 2515 Speedway, Austin, TX 78712-1192, USA  }

${}^{2}${\small \sl 
Walter Burke Institute for Theoretical Physics \\
California Institute of Technology, Pasadena, CA 91125, USA
}

${}^{3}${\small \sl RIKEN Interdisciplinary Theoretical and Mathematical Sciences (iTHEMS), \\Wako, Saitama 351-0198, Japan}

${}^4$ {\small \sl Kavli Institute for the Physics and Mathematics of the Universe (WPI) \\
University of Tokyo, Kashiwa 277-8583, Japan}

\vskip 2em
\end{center}

\begin{abstract}

When an interface connects two CFTs, the entanglement entropy between the two CFTs is determined by a quantity called the effective central charge. The effective central charge does not have a simple form in terms of the central charges of the two CFTs, but intricately depends on the transmissive properties of the interface.

In this article, we examine universal properties of the effective central charge. We first clarify how the effective central charge appears when considering general subsystems of the interface CFT. Then using this result and ideas used in the proof of the $c$-theorem, we provide a universal upper bound on the effective central charge.

In past studies, the effective central charge was defined only in two dimensions.
We propose an analogue of the effective central charge in general dimensions possessing similar universal properties as in two dimensions.

 \end{abstract}

\end{titlepage}

\restoregeometry

\tableofcontents

\section{Introduction}
\label{sec:intro}

Entanglement entropy is an important tool for capturing the global structure of a given quantum system.
Given a subsystem $A$, this quantity can be evaluated by using the replica trick,
\begin{equation}
S_A \equiv -\tr \rho_A \ln \rho_A = \lim_{n \to 1} S_A^{(n)},
\end{equation}
where $S_A^{(n)}$ is the R\'enyi entropy of order $n$ defined as
\begin{equation}
S_A^{(n)} = \fr{1}{1-n} \ln \tr \rho_A^n.
\end{equation}
Here, the reduced density matrix $\rho_A$ is obtained by tracing out the complement of the subsystem $A$. 
Due to its usefulness, this quantity has been studied in various situations.
In this article, we are particularly interested in the entanglement entropy in interface CFTs (ICFTs),
which constitute a class of CFTs where two (possibly different) CFTs are connected along a codimension-$1$ interface \cite{Oshikawa1996,Oshikawa1997,Bachas2002}.
The entanglement entropy in ICFTs has been extensively investigated in various (1+1)-$d$ CFT setups \cite{Sakai2008, Brehm2015,Brehm2016,Wen2018,Gutperle2016a}.

\begin{figure}[t]
 \begin{center}
  \includegraphics[width=15.0cm,clip]{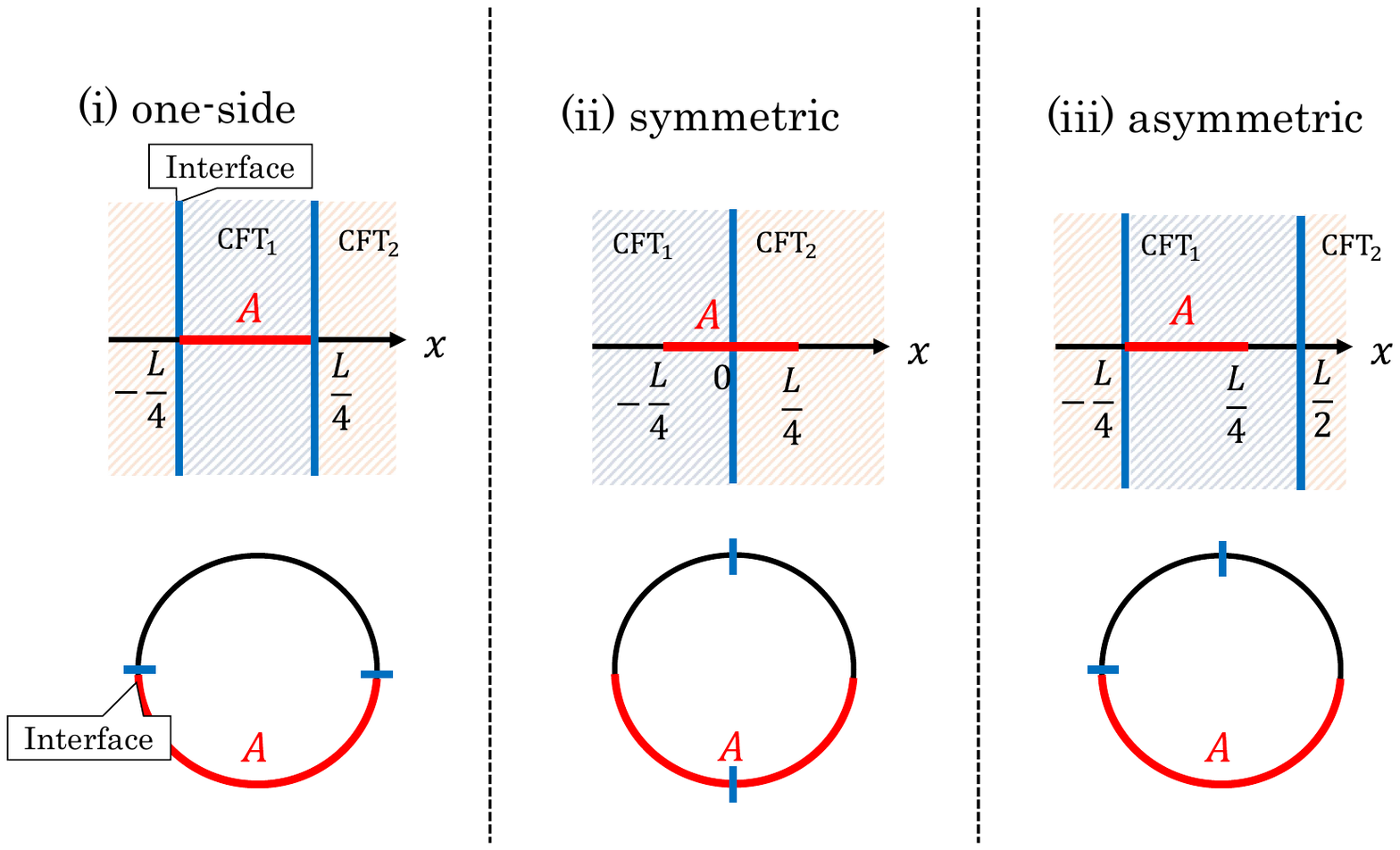}
 \end{center}
 \caption{Setups of our interest.
 The red line represents the branch cut in the subsystem $A$ and the blue line represents the interface connecting CFT${}_1$ and CFT${}_2$.
 The bottom picture shows the time slice of each system.
 }
 \label{fig:setup}
\end{figure}

In this article, we consider the entanglement entropy of the ground state of a given ICFT (constituted by CFT${}_1$ and CFT${}_2$) on a finite spatially periodic system of size $L$.
There are various choices for the location of the interface.
We particularly focus on the following three setups:
\begin{enumerate}[(i)]

\item the subsystem $A$ is located on one side of the interface, $A=[-L/4,L/4]$ with interfaces located at $x=\pm L/4$ (see the left of Figure \ref{fig:setup}),

\item the subsystem $A$ is set to be $A= [-L/4,L/4]$ containing the interface located at $x=0$ (see the center of Figure \ref{fig:setup}),

\item the subsystem $A$ is set to be $A=[-L/4,L/4]$ and the interfaces are inserted at $x=-L/4$ and $x=L/2$, respectively (see the right of Figure \ref{fig:setup}).

\end{enumerate}
While we choose the subsystem $A=[-L/4,L/4]$ for simplicity,
we can generalize it to $A=[-a,a]$, which does not change the logarithmic divergent term of the entanglement entropy but gives a finite correction.
We will discuss this later in Section \ref{sec:gen}.

The result for each setup is as follows:
\begin{enumerate}[Setup (i)]

\item
\begin{equation}\label{eq:resulti}
S_A = \fr{c_{\text{eff}}}{3}\ln \fr{L}{\pi \epsilon} + \text{const}.
\end{equation}

\item
\begin{equation}
S_A = \fr{c_1+c_2}{6} \ln \fr{L}{\pi \epsilon} + \text{const}.
\end{equation}

\item
\begin{equation}\label{eq:resultiii}
S_A = \fr{c_{\text{eff}}+c_1}{6} \ln \fr{L}{\pi \epsilon} + \text{const}.
\end{equation}

\end{enumerate}
Throughout this article, "const." means a constant with respect to $\epsilon$.
The first setup has been explored in various works, such as \cite{Sakai2008, Brehm2015}.
The coefficient of the logarithmically divergent term is given not only by a simple form in terms of $c_1$ and $c_2$,
but it also involves a quantity $c_{\text{eff}}$ called the {\it effective central charge}, which depends on the details of the interface between CFT${}_1$ and CFT${}_2$.
For example, 
while $c_{\text{eff}}$ is given by $c_1$ ($=c_2$) for totally transmissive interfaces,
$c_{\text{eff}}$ vanishes for totally reflective interfaces.
There is no universal form for partially transmissive interfaces.
It is natural to ask what is the physical meaning of $c_{\text{eff}}$.
To answer this question,
we will explore the universal property of $c_{\text{eff}}$ in this article.

It is natural to expect that the entanglement entropy in setup (iii) has the simple form (\ref{eq:resultiii}).
However, it is non-trivial that $c_{\text{eff}}$ in (\ref{eq:resultiii}) is really the same as the effective central charge that appears in (\ref{eq:resulti}).
In fact, this is true in holographic CFTs \cite{Karch2021,Karch:2022vot},
where the entanglement entropy is given by the area of the Ryu-Takayanagi surface through the AdS/ICFT correspondence. The relation also holds in some free field theory examples \cite{Kruthoff:2021vgv}.
Can we show this universal formula in general CFTs?
One can immediately find that the well-known method to evaluate the entanglement entropy in the presence of interfaces cannot be applied to setup (iii), as discussed in \cite{Karch:2022vot}.
In this article, we will introduce a method to overcome this problem and show a universal bound for $c_{\text{eff}}$ in {\it general} (1+1)-$d$ ICFTs.

The relation between entanglement entropy and central charge in higher dimension is always of deep interest as illustrated in e.g., \cite{Myers:2010xs,Myers:2010tj}, and it is also explored here. More specifically, we are able to generalize the above discussion to the entanglement entropy of codimension-1 subsystems in higher-dimensional ICFTs. In higher dimensions, the shape (or topology) of the entangling surface (and the spacetime geometry) matters.\footnote{We know from \cite{Solodukhin:2008dh,Fursaev:2012mp,Liu:2012eea} that already at $d=4$, the universal log-divergent term $s_{(4)}^{\text{uni}}$ in the entanglement entropy depends on the shape of the entangling surface $\Sigma$ as $$s_{(4)}^{\text{uni}}=-2a_4\int_{\Sigma}\sqrt{h}\frac{\mathcal{R}}{4\pi}-c_4\int_{\Sigma}\sqrt{h}\frac{1}{2\pi}\left(\frac{K^2}{2}-K_{ab}K^{ab}\right),$$ where $\mathcal{R}$ is the intrinsic curvature on $\Sigma$, and $K_{ab}$ is the extrinsic curvature for $\Sigma$. $a_4$ is the coefficient of the Euler density term in $d=4$ Weyl anomaly, and $c_4$ is that of the Weyl tensor term. In our spherical geometry case $s_{(4)}^{\text{uni}}=-4a_4$, whereas in e.g. cylindrical geometry case, we have $s_{(4)}^{\text{uni}}=-\frac{c_4}{2}\frac{L}{R}$ ($L$ and $R$ are the length and radius of the cylinder). Notice that below we will use $c_1$ and $c_2$ to denote those A-type coefficients for CFT$_1$ and CFT$_2$, to keep notations consistent with the $d=2$ case. \label{foot:typec}} For the purpose of this work, we restrict ourselves to $d$-dimensional ICFTs with spacetime geometry $\mathbb{R}\times S^{d-1}$ with the interface along the equator (with geometry $\mathbb{R}\times S^{d-2}$), and focus on subsystems extending away from the interface. In other words, we only consider spherical entangling surfaces $S^{d-2}$ parallel to or overlapping with the interface. By introducing a UV cutoff $\epsilon$, we can extract a scheme-independent (universal) logarithmically divergent term (for even-dimensional ICFTs) or constant term (for odd-dimensional ICFTs) in the entanglement entropy of subsystem $A$, thus defining a similar $c_{\text{eff}}$ as in \eqref{eq:resulti} for the subsystem that is the entire spacelike region on one side of the interface. Namely, for the half-space entanglement entropy for a $d$-dimensional ICFT ($d>2$), that is an entangling surface coincident with the interface, we define 
\begin{equation}
    \begin{split}
        S_{EE}^{\text{(i)}}&=\dots+(-1)^{\frac{d}{2}-1}4c_{\text{eff}}~\log\frac{2}{\epsilon}\  +\mathrm{const.},\quad \mathrm{even}\  d,\\ 
        S_{EE}^{\text{(i)}}&=\dots+(-1)^{\frac{d-1}{2}}2\pi~ c_{\text{eff}},\quad \mathrm{odd}\  d.\\
    \end{split}
    \label{eq:higherceff}
\end{equation}

We will also see below that for case (ii) where the subsystem surrounds the interface (two entangling surfaces parallel to the interface on either side), the universal terms in even dimensions are given by
\begin{equation}
    \begin{split}
        S_{EE}^{\text{(ii)}}&=\dots+(-1)^{\frac{d}{2}-1}2(c_1+c_2)~\log\frac{2}{\epsilon}\  +\mathrm{const.},\quad \mathrm{even}\  d,\\ 
    \end{split}
    \label{eq:dresultii}
\end{equation}

 In even dimensions, the coefficients $c_1$, $c_2$ are the A-type trace anomaly (coefficient of the Euler density term) of CFT$_1$ and CFT$_2$ on spheres, which is the geometry we adopt in this paper \cite{Myers:2010xs,Myers:2010tj,Casini:2011kv}. However, for odd dimensions, there is no Weyl anomaly. Alas, we can still define "central charges" $c_1$, $c_2$ for CFT$_1$ and CFT$_2$ via the Weyl anomaly coefficients of a boundary even-dimensional theory\cite{Herzog:2013ed}, or via the short distance behavior of stress tensor correlators inside each of the CFTs. Note that in all cases, $c_1$, $c_2$ are defined independently from the holographic entanglement entropy. In Section \ref{sec:higherholo} we will give an expression for them in terms of the bulk warpfactor $A$ for both even and odd dimensions following \cite{Myers:2010tj}.

Universal relations similar to \eqref{eq:resultiii} have recently been found in four-dimensional holographic ICFTs \cite{Uhlemann:2023oea} for setup (iii), and in this paper we will prove the generalized version for any even dimension:
\begin{equation}
    \begin{split}
        S_{EE}^{\text{(iii)}}&=\dots+(-1)^{\frac{d}{2}-1}2(c_1+c_{\text{eff}})~\log\frac{2}{\epsilon}\  +\mathrm{const.},\quad \mathrm{even}\  d,\\ 
    \end{split}
    \label{eq:dresultiii}
\end{equation}

Notice that we do not give the universal constant terms for case (ii) and (iii) in odd dimensions, for reasons that will be explained in Section \ref{sec:higherholo}.

One interesting application of (\ref{eq:resulti}) - (\ref{eq:resultiii}) is to show a universal bound on the effective central charge,
\begin{equation}
\label{eq:bound}
c_{\text{eff}} \leq \min(c_1, c_2).
\end{equation}
While this bound is naturally expected from the quasiparticle picture (see \cite{Wen2018} and Section \ref{sec:beyond}),
heretofore there is no proof of it.
Against this backdrop,
another aim of this article is to show this bound in general ICFTs.
One well-known approach to give an inequality for central charges is used in the entropic proof of the $c$-theorem \cite{Casini2007}.
In this article, we will see that a generalization of the entropic $c$-theorem to ICFTs can be used to show the inequality (\ref{eq:bound}), also using our results (\ref{eq:resulti}) - (\ref{eq:resultiii}).
We will also give an alternative proof of the bound (\ref{eq:bound}) using the AdS/ICFT correspondence in Section \ref{sec:bound}. Furthermore, in Section \ref{sec:higherholo} we will show that in even and odd dimensional holographic ICFT on a sphere, this bound still holds for $c_{\text{eff}}$ and the Weyl anomalies $c_1,c_2$, providing a holographic proof. 

Finally in Section \ref{sec:disc}, we will discuss possible Casini--Huerta style \cite{Casini2007} proof of \eqref{eq:bound} in general dimensions, among other interesting open problems.

\section{Universal Entanglement Entropy in ICFTs}\label{sec:uni}

In this section, we will present a CFT analysis of the entanglement entropy in the presence of interfaces.
We assume for the interfaces to be conformal interfaces, which satisfy the following condition,
\begin{equation}\label{eq:I}
\pa{L_n^{(1)} - \bar{L}_{-n}^{(1)} }I = I \pa{L_n^{(2)} - \bar{L}_{-n}^{(2)} }  \ \ \ \ \ \text{for any } n,
\end{equation}
where $L_n^{(i)}$ and $\bar{L}_{n}^{(i)}$ are the holomorphic and anti-holomorphic Virasoro algebra of CFT${}_i$.
Most studies on conformal interfaces have focused on a special class of interfaces called topological interfaces,
which greatly simplify CFT calculations due to the commutativity of the Viraosro generators and topological interface.
On the other hand, dealing with general conformal interfaces is difficult due to the lack of symmetry.
Nonetheless, some quantities can still be calculated using the remaining symmetry.
In this article, we will consider general conformal interfaces.

\subsection{Setup (i)}

\begin{figure}[t]
 \begin{center}
  \includegraphics[width=5.0cm,clip]{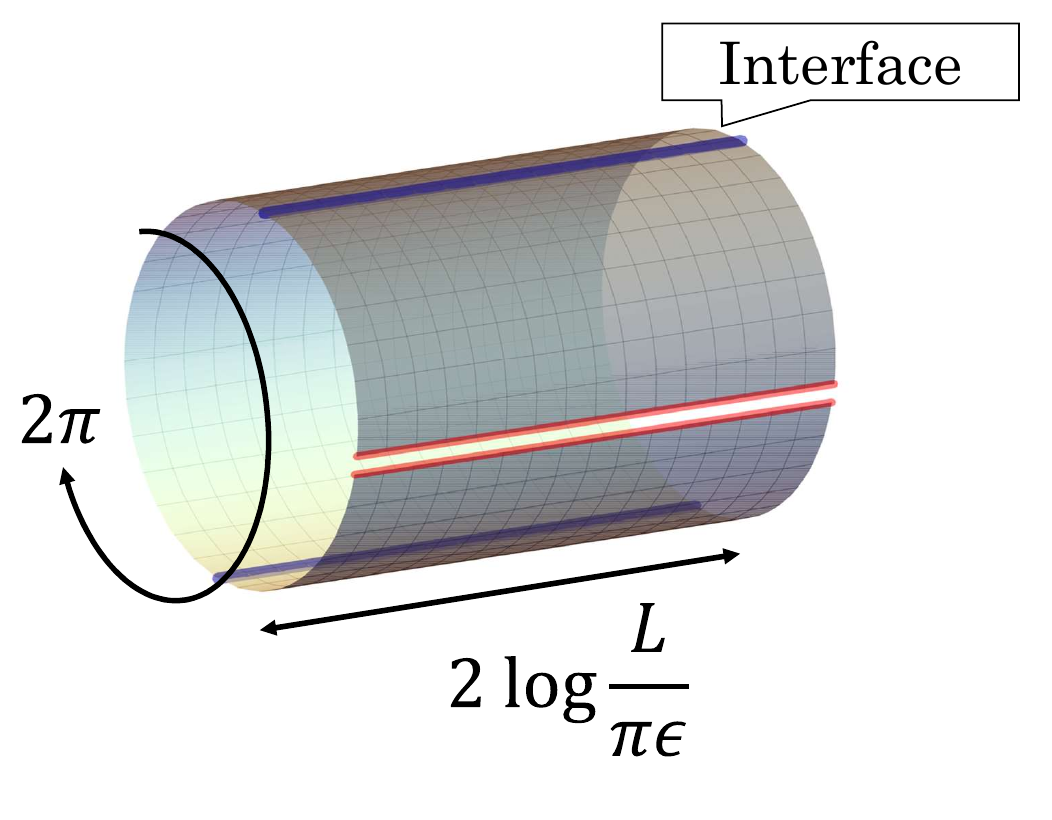}
 \end{center}
 \caption{The cylinder obtained from a single replica sheet by the conformal maps (a)$\sim$(c).
  The red line is the branch cut in the subsystem $A$ and the blue line corresponds to the interface connecting CFT${}_1$ and CFT${}_2$.}
 \label{fig:cylinder}
\end{figure}

Setup (i) has been studied extensively in previous works, such as \cite{Sakai2008, Brehm2015}.
The entanglement entropy for setup (i) has the following universal form,
\begin{equation}
S_A = \fr{c_{\text{eff}}}{3}\ln \fr{L}{\pi \epsilon} + \text{const}.
\end{equation}
where we factorize the CFT Hilbert space by cutting out a disk-shaped hole of radius $\epsilon \ll L$.
In other words, $\epsilon$ is the UV cutoff parameter.
The effective central charge is defined in \cite{Gutperle2016}.
As we will see below,
this effective central charge is closely related to the vacuum energy of the interface Hilbert space.
To see this, it is convenient to consider the following conformal maps:
\begin{enumerate}[(a)]

\item Mapping from cylinder to sphere: $z \to \ex{\fr{2\pi i }{L}z}$.

\item Rotation: $z \to \fr{z+1}{z-1} $.

\item Mapping from sphere to cylinder $z \to \ln z$.

\end{enumerate}
Using these maps, each replica sheet is mapped to a cylinder illustrated in Figure \ref{fig:cylinder}.
The R\'enyi entropy is now expressed in terms of the closed string amplitude as
\begin{equation}
S_A^{(n)} = \fr{1}{1-n} \ln \fr{Z_n}{(Z_1)^n},
\end{equation}
where the amplitude is defined by
\begin{equation}
Z_n = \bra{\ti{b}_n} \ex{-\fr{1}{2\pi n} 2 \ln \fr{L}{\pi \epsilon} H^{I}_n } \ket{\ti{b}_n}.
\end{equation}
The boundary state $\ket{\ti{b}_n}$ represents a physical boundary condition imposed on the entangling surface (see its detailed discussion in \cite{Ohmori2015}).
\footnote{
Here we implicitly assume that the vacuum couples to the boundary $\ket{\ti{b}_n}$.
}
The Hamiltonian $H^I_n$ describes the Hamiltonian for the closed string.
Since in the limit $\epsilon \ll L$ the length of the cylinder is very long,
only the lowest energy state propagates through the cylinder.
Note that the interface Hilbert space $\ca{H}^I_n$ can be defined using modular invariance as
\begin{equation}\label{eq:HI}
\tr_{\ca{H}^I_n }  q^{\fr{1}{n}H^I_n } = \tr  \pa{\ti{q}^{H_1} I \ti{q}^{H_2} I^\dagger}^n,
\end{equation}
where $q=\ex{2\pi i \tau}$ and $\ti{q} = \ex{-2\pi i \fr{1}{\tau}}$.
Here, we label the conformal interface by $I$.
The Hamiltonian $H_i$ describes the Hamiltonian of the open string on CFT${}_i$.
Let us denote the lowest conformal dimension in the interface Hilbert space by $\Delta^0_n$.
Then the R\'enyi entropy has the following form,
\begin{equation}
S_A = \fr{c_{\text{eff}}}{3}  \ln \fr{L}{\pi \epsilon} + 2 \ln g^I,
\end{equation}
where the effective central charge is defined by
\begin{equation}\label{eq:ceff}
c_\text{eff} \equiv \lim_{n \to 1} \fr{12n}{1-n^2} \pa{n\Delta^0_1 - \fr{\Delta^0_n}{n}},
\end{equation}
and the boundary entropy is defined by
\begin{equation}
\ln g^I \equiv \lim_{n \to 1} \fr{1}{1-n} \pa{\ln \braket{\ti{b}_n | \Delta^0_n   } - n \ln \braket{\ti{b}_1 | \Delta^0_1   } }.
\end{equation}
Note that the effective central charge (\ref{eq:ceff}) reduces to the central charge of the CFT if the interface is totally transmissive.
On the other hand, $c_{\text{eff}}$ vanishes if the interface is totally reflective, as in the case of a BCFT.

\subsection{Setup (ii)}

In a similar manner, the entanglement entropy for the setup (ii) can be evaluated.
We begin by applying the same conformal transformations to the replica manifold corresponding to (ii),
following which the corresponding closed string amplitude is given by
\begin{equation}
Z_n = \bra{\ti{b}^{(1)}} \ex{-\fr{1}{2\pi n} \ln \fr{L}{\pi \epsilon} H_1  }   I \ex{-\fr{1}{2\pi n} \ln \fr{L}{\pi \epsilon} H_2  }     \ket{\ti{b}^{(2)}}.
\end{equation}
In this case, we cannot impose the same boundary condition on two entangling surfaces,
and thus we label the two different boundaries by $\ti{b}^{(i)}$, which corresponds to a boundary state in CFT${}_i$.
We assume that the vacuum in CFT${}_1$ couples to the vacuum in CFT${}_2$ through the interface.
Note that this coupling constant can be interpreted as the boundary entropy by using the folding trick.
We will denote this boundary entropy by $\ln g_{1|2}$.
In the $\epsilon \ll L$ limit,
on one side of the interface, only the lowest energy state in $\ca{H}_1$ propagates
and on the other side, only the lowest energy state in $\ca{H}_2$ propagates.
Therefore, the amplitude can be approximated by
\begin{equation}
Z_n = \bra{\ti{b}^{(1)}} \ex{\fr{1}{ n} \ln \fr{L}{\pi \epsilon} \fr{c_1}{12}  }   I \ex{\fr{1}{ n} \ln \fr{L}{\pi \epsilon} \fr{c_2}{12}}     \ket{\ti{b}^{(2)}}.
\end{equation}
Then we obtain the entanglement entropy as
\begin{equation}\label{eq:EEii}
S_A = \fr{c_1+c_2}{6} \ln \fr{L}{\pi \epsilon} + \ln g_1 + \ln g_2 + \ln g_{1|2},
\end{equation}
where we define $g_i \equiv \braket{\ti{b}^{(i)} | 0}    $.
Using the folding trick,
this result can be interpreted as the entanglement entropy for the vacuum state in a product CFT, $\text{CFT}_1 \otimes \text{CFT}_2$, with a conformal boundary.

\subsection{Setup (iii)}
While the general form of the entanglement entropy for the setup (iii) has not been derived,
the universal formula has been conjectured in \cite{Karch2021,Karch:2022vot},
\begin{equation}\label{eq:conj}
S_A = \fr{c_{\text{eff}} + c_1}{6} \ln \fr{L}{\pi \epsilon} + \text{const}.
\end{equation}
In this section, we will verify this conjecture using CFT analysis.

\begin{figure}[t]
 \begin{center}
  \includegraphics[width=5.0cm,clip]{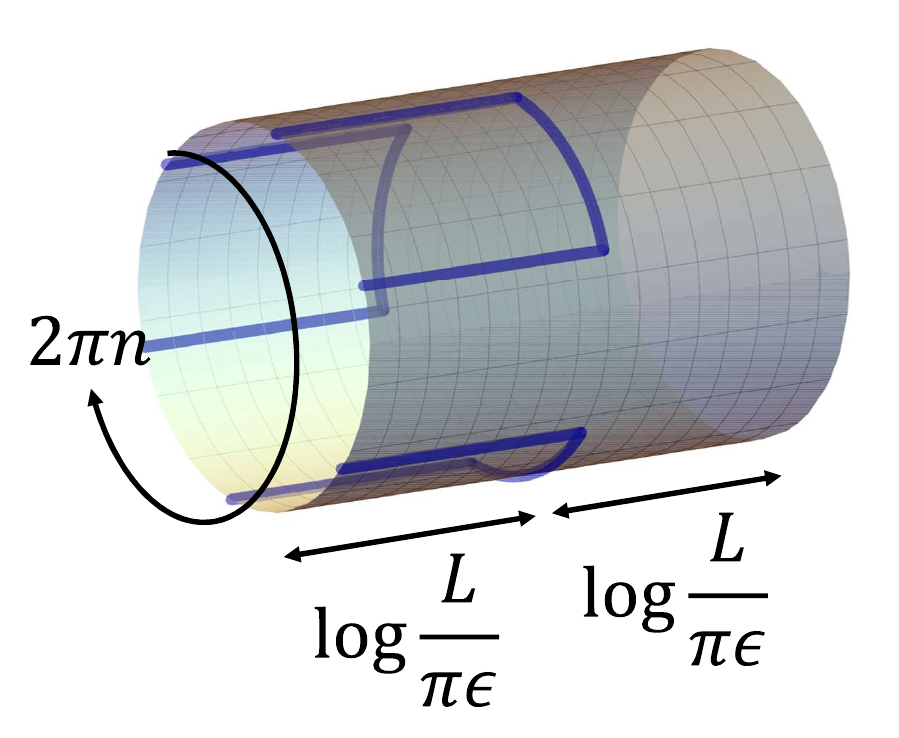}
    \includegraphics[width=5.0cm,clip]{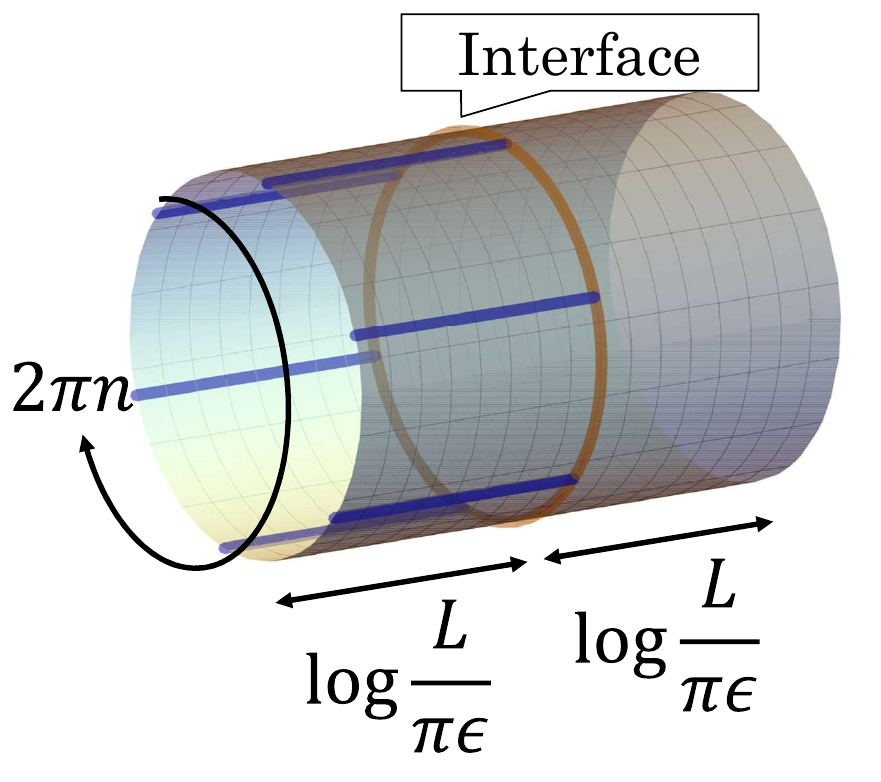}
 \end{center}
 \caption{(Left) The closed string amplitude corresponding to the replica manifold for the setup (iii).
 The blue line represents the interface connecting CFT${}_1$ and CFT${}_2$.
 (Right) This amplitude can be thought of as an amplitude connecting the interface CFT (\ref{eq:HI}) to the CFT${}_1$ through a certain interface  $I'$.
 }
 \label{fig:asym}
\end{figure}

Let us consider the conformal map (a)$\sim$(c) for the replica manifold related to the setup (iii).
The closed string amplitude that we obtain can be depicted as the left of Figure \ref{fig:asym}.
This amplitude can be thought of as an amplitude with an interface between the interface Hilbert space $\ca{H}^I_n$ (see equation (\ref{eq:HI})) and $\ca{H}_1$.
Therefore, this amplitude can be expressed as
\begin{equation}\label{eq:closediii}
Z_n = \bra{\ti{b}^{(1)}} \ex{-\fr{1}{2\pi n} \ln \fr{L}{\pi \epsilon} H^I_n  }   I' \ex{-\fr{1}{2\pi n} \ln \fr{L}{\pi \epsilon} H_1  }     \ket{\ti{b}^{(2)}},
\end{equation}
where the interface $I'$ is different from the interface $I$. This interpretation is depicted on the right of Figure \ref{fig:asym}.
The key point is that in the $\epsilon \ll L$ limit, on one side of the interface, only the lowest energy state in $\ca{H}^I_n$ propagates
and on the other side, only the lowest energy state in $\ca{H}_1$ propagates.
This allows for a straightforward calculation of the amplitude, similar to that for the setup (ii). 
It should be noted that here we assume that the vacuum in $\ca{H}^I_n$ couples to the vacuum in $\ca{H}_1$ through the interface $I'$.
Consequently, we obtain the entanglement entropy as
\begin{equation}\label{eq:EEiii}
S_A = \fr{c_{\text{eff}}+c_1}{6} \ln \fr{L}{\pi \epsilon} + \ln g_1 + \ln g_I + \ln g_{I|1},
\end{equation}
where $\ln g_{I|1}$ is the boundary entropy defined by the folding trick.
This completely matches the conjecture (\ref{eq:conj}).
This is one of the main results of this article.
It might be possible to express $\ln g_{I|1}$ in terms of the boundary entropy of the interface $I$ by careful consideration of the interface $I'$.
We leave this analysis for further work.

\subsection{Generalization}
\label{sec:gen}

\begin{figure}[t]
 \begin{center}
  \includegraphics[width=15.0cm,clip]{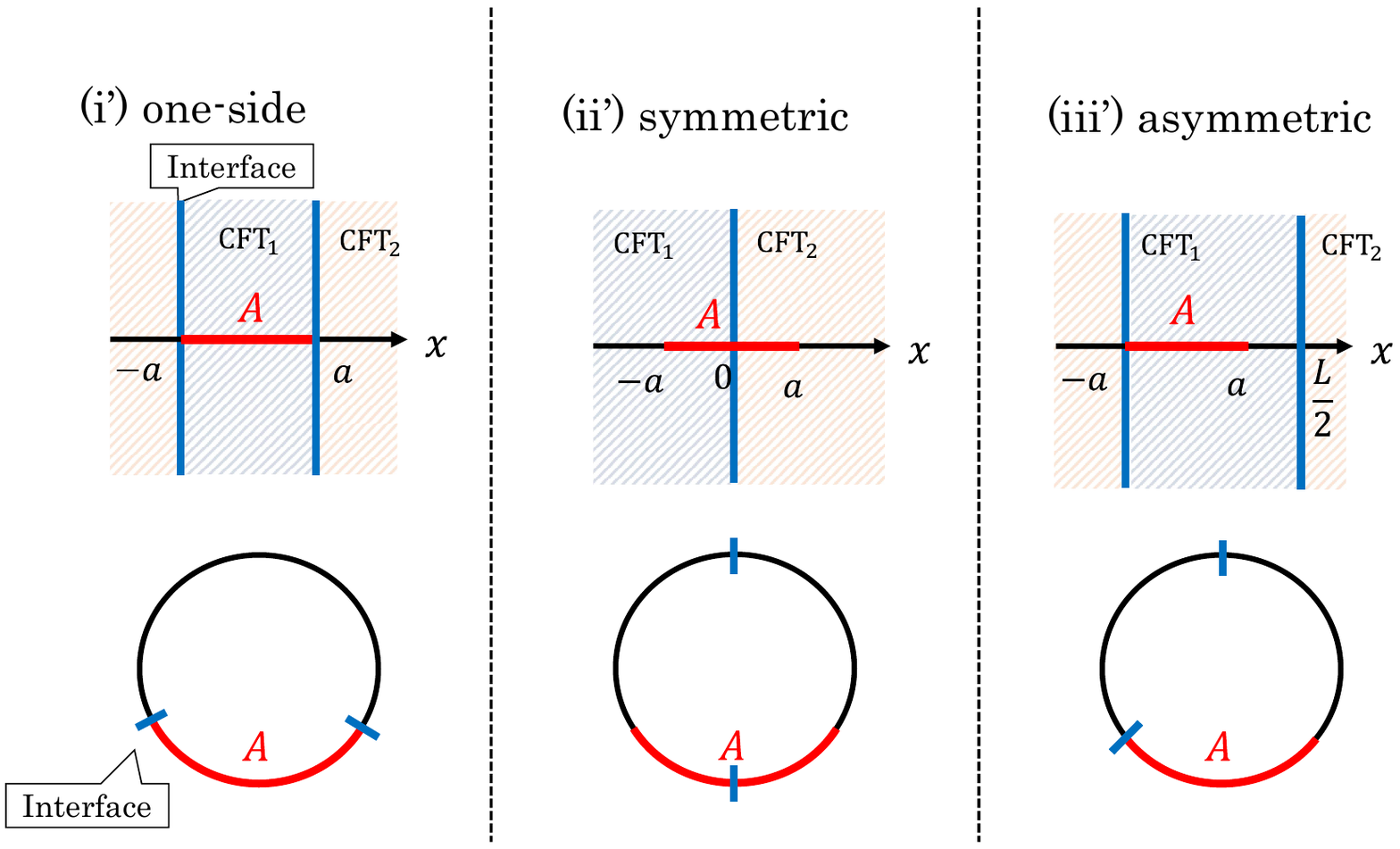}
 \end{center}
 \caption{A slight generalization of our setups in Figure \ref{fig:setup}.}
 \label{fig:setup2}
\end{figure}

\begin{figure}[t]
 \begin{center}
  \includegraphics[width=5.0cm,clip]{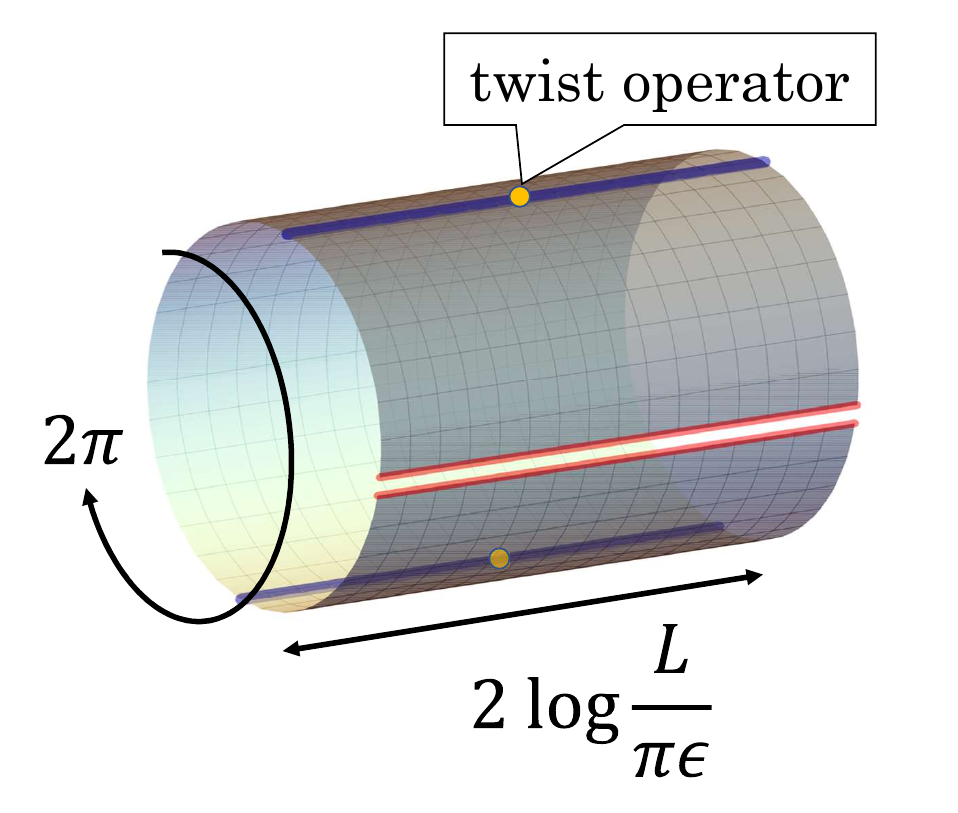}
 \end{center}
 \caption{The cylinder obtained from one replica sheet by the conformal map.
The change in the subsystem size causes a slight deformation of the cylinder near $t=0$. 
 Nevertheless, the lowest energy approximation that was used to evaluate the closed string amplitude remains valid.
 }
 \label{fig:twist}
\end{figure}

In this section, we provide comments on the entanglement entropy for slightly generalized setups.
In the previous sections, we focused only on a specific subsystem choice $A=[-L/4, L/4]$.
We now generalize it to $A=[-a,a]$ (see Figure \ref{fig:setup2}).
Let us first consider setup (i').
For convenience, we rescale the cylinder by the conformal map
\begin{equation}
z \to z^{\fr{L}{2a}}.
\end{equation}
Then the size of the subsystem $A$ becomes the same as that of its complement.
Note that this conformal map introduces a twist operator on the interface.
Applying the conformal map (a)$\sim$(c),
we obtain the closed string amplitude as shown in Figure \ref{fig:twist}.
The difference between (i) and its generalization (i') is that
the twist operators are inserted in the replica manifold for the setup (i').
This operator insertion plays a similar role in the interface.
That is, this gives a finite correction to the entanglement entropy as
\begin{equation}
S_A = \fr{c_{\text{eff}}}{3}  \ln \pa{ \fr{L}{\pi \epsilon}  \sin \pa{  \fr{2\pi a  }{L} }}  + \text{const}.
\end{equation}
In other words, the coefficient of the logarithmic divergent term is still universal and given by the effective central charge.

The same logic can be applied to the setup (iii') and we obtain
\begin{equation}
S_A = \fr{c_{\text{eff}}+c_1}{6} \ln \pa{ \fr{L}{\pi \epsilon}  \sin \pa{  \fr{2\pi a  }{L} }} + \text{const}.
\end{equation}
The main point is that despite the deformation of the interface $I'$ caused by the variation of the subsystem size,
the lowest energy approximation that was used to derive (\ref{eq:EEiii}) from (\ref{eq:closediii}) is still valid.
Therefore, the coefficient of the logarithmic divergent part remains unchanged.

The setup (ii') is easily addressed by just the folding trick and we obtain
\begin{equation}
S_A = \fr{c_1 + c_2}{6}  \ln \pa{ \fr{L}{\pi \epsilon}  \sin \pa{  \fr{2\pi a  }{L} }}  + \text{const}.
\end{equation}
In this analysis, it is impossible to shed light on the constant term.
We leave this for future work.

\section{Beyond Vacuum State}\label{sec:beyond}

\begin{figure}[t]
 \begin{center}
  \includegraphics[width=15.0cm,clip]{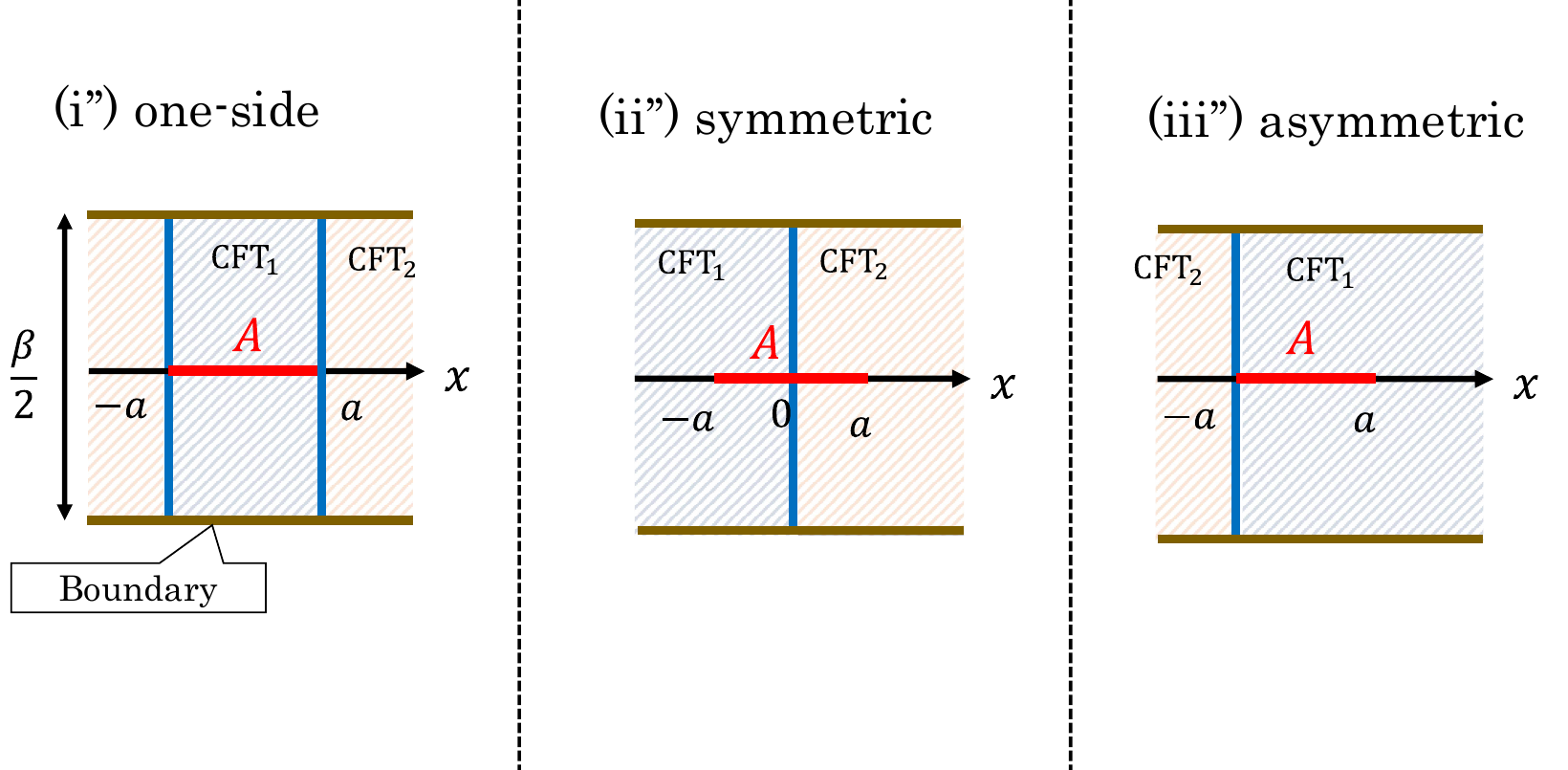}
 \end{center}
 \caption{Global quenched setups of our interest.
 These setups resemble the setups shown in Figure \ref{fig:setup2}.
 The difference is that our system is located on an infinite line and is excited by a global quench.
 The brown lines represent the boundaries, which are located along $\tau=\beta/4$ and $\tau=-\beta/4$.
 }
 \label{fig:setup3}
\end{figure}

It would be interesting to investigate whether the formula (\ref{eq:conj}) remains applicable to non-equilibrium systems.
An interesting and accessible system to study can be obtained by a global quantum quench.
The system is prepared in the ground state of a gapped Hamiltonian $H_0$.
When we instantaneously change the Hamiltonian to $H$,
the state is highly excited with respect to the new Hamiltonian $H$.
In a CFT, we can prepare this global quenched state by smearing a boundary state $\ket{\ti{B}}$ as \cite{Calabrese2005,Calabrese2007}
\begin{equation}
\ket{\Psi(0)} = \ex{-\beta \fr{H}{4}} \ket{\ti{B}},
\end{equation}
where $\beta$ is the effective temperature and the high-temperature limit ($\beta\ll L$) results in a system with only short-ranged entanglement.
The time evolution of the system is given by
\begin{equation}
\ket{\Psi(t)} = \ex{-iHt}  \ket{\Psi(0)}.
\end{equation}
Let us consider a CFT on an infinite line.
Our focus is on the entanglement entropy for the global quenched state with an interface inserted. 
We are particularly interested in the following three setups, which resemble those considered in the previous sections.
\begin{enumerate}[(i'')]

\item the subsystem $A$ is on one side of the interface, $A=[-a,a]$ (see the left of Figure \ref{fig:setup3}),

\item the subsystem $A$ is set to be $A= [-a,a]$ and the interface is located at $x=0$ (see the center of Figure \ref{fig:setup3}),

\item the subsystem $A$ is set to be $A=[-a,a]$ and the interface is located at $x=-a$ (see the right of Figure \ref{fig:setup3}).

\end{enumerate}

While these setups are significantly more complicated than the setups in the vacuum state,
it is still possible to analytically evaluate the entanglement entropy in this case. 
The point is that the high-temperature limit, which now means $\beta \ll a$, greatly simplifies the CFT calculation.
Our goal is to identify the universal behavior of the entanglement dynamics,
regardless of the boundary state's specific details.
To achieve this objective, 
we can use a slightly different quench model introduced  in \cite{Hartman2013}, called the thermofield double (TFD) state,
\begin{equation}
\ket{\Psi(t)} = \sum_n  \ex{-iHt-H\fr{\beta}{2}} \ket{n}_1 \otimes \ket{n}_2,
\end{equation}
where $\ket{n}_i$ is an eigenstate in CFT${}_i$.
This strategy for simplification has been utilized in previous studies, such as \cite{Asplund2015, KudlerFlam2020,Kudler-Flam:2022irq}.
With this simplification, each sheet is on a cylinder with a circumference of $\beta$.

\begin{figure}[t]
 \begin{center}
  \includegraphics[width=13.0cm,clip]{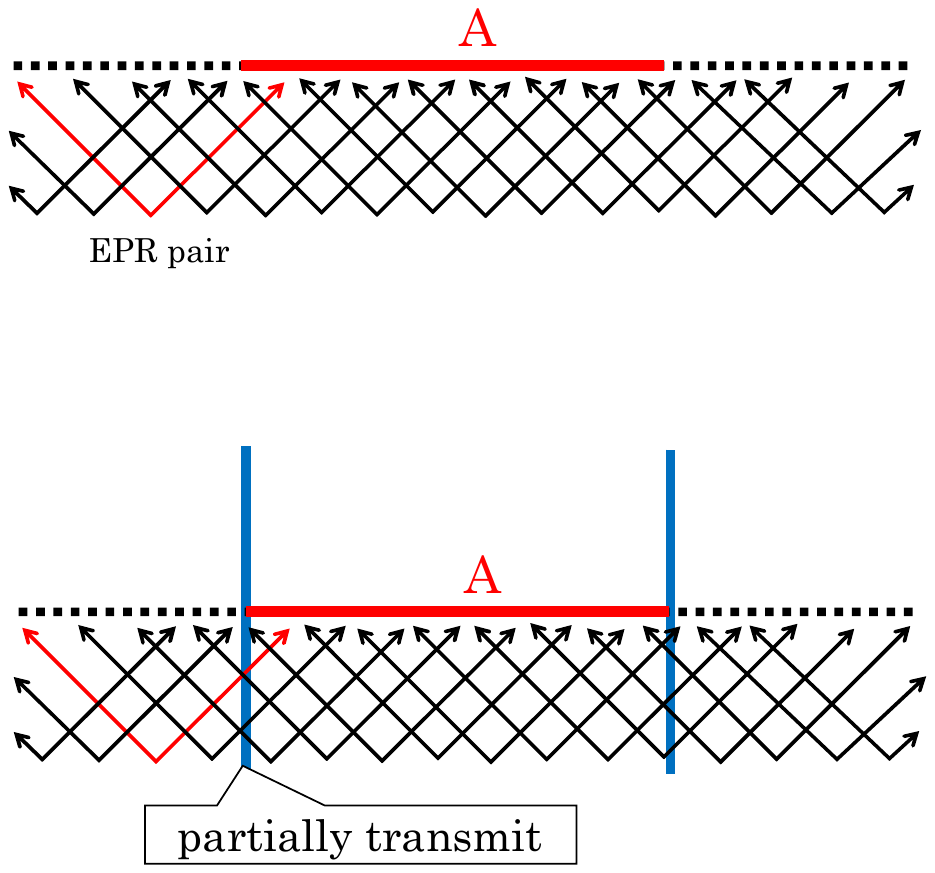}
 \end{center}
 \caption{(Top) Quasiparticle picture for the entanglement entropy.
 The arrows represent the quasiparticle picture of the information spreading under the global quench.
 (Bottom) This picture explains why the entanglement entropy decreases when a partially transmissive interface is present. 
Only a portion of the quasiparticles can transmit through the interface.
 }
 \label{fig:epr}
\end{figure}

Without the interface, the entanglement entropy is calculated as follows,
\begin{equation}\label{eq:global2}
\begin{aligned}
S_A(t) =\left\{
    \begin{array}{ll}
     \fr{4\pi c}{3 \beta} t + \fr{2c}{3} \log \fr{\beta}{2\pi \epsilon}  ,& \text{if }  0<t<a ,\\
     \fr{4\pi c}{3 \beta} a + \fr{2c}{3} \log \fr{\beta}{2\pi \epsilon}    ,& \text{if }  a<t .\\
    \end{array}
  \right.\\
\end{aligned}
\end{equation}
This result can be interpreted as counting the EPR (Einstein-Podolsky-Rosen) pairs created by the global quench, as depicted in the top panel of Figure \ref{fig:epr}. 
The EPR pairs are created at $t=0$ and propagate at the speed of light.
When one of the partners in the pair enters the subsystem $A$ and the other remains in its complement, the entanglement entropy increases.
Once the system is fully thermalized at $t=a$, the entanglement entropy becomes constant.
From the expression (\ref{eq:global2}),
an analogue of the formula (\ref{eq:conj}) may be given by
\begin{equation}
\begin{aligned}
S_A(t) =\left\{
    \begin{array}{ll}
     \fr{c_{\text{eff}}+c_1}{6}  \fr{4\pi }{ \beta} t + \fr{c_{\text{eff}}+c_1}{3} \log \fr{\beta}{2\pi \epsilon}  ,& \text{if }  0<t<t_* ,\\
   \fr{c_1}{3}    \fr{4\pi }{ \beta} a + \fr{c_{\text{eff}}+c_1}{3} \log \fr{\beta}{2\pi \epsilon}    ,& \text{if }  t_*<t ,\\
    \end{array}
  \right.\\
\end{aligned}
\end{equation}
where $t_*=\fr{2c_1}{c_1+c_{\text{eff}}}$.
We verify this conjecture in this section.

\subsection{Setup (i'')}
We begin by considering the setup (i'').
To provide a simple explanation,
we introduce the twist operator on the interface.
The replica partition function can then be described in terms of the twisted operator.
In the high-temperature limit, this four-point function of twist operators factorizes into a product of two two-point functions,
which can be illustrated as
\newsavebox{\boxaa}
\sbox{\boxaa}{\includegraphics[width=300pt]{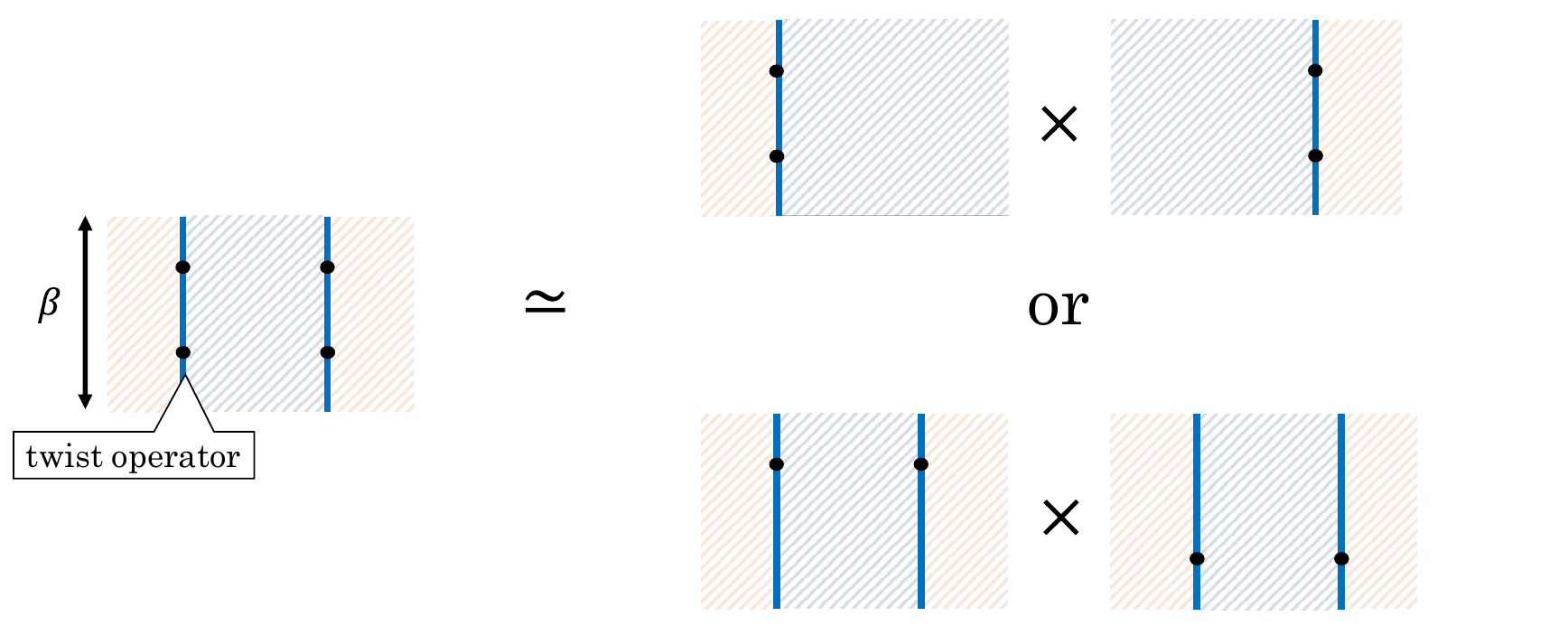}}
\newlength{\boxwaa}
\settowidth{\boxwaa}{\usebox{\boxaa}} 

\begin{equation}
\parbox{\boxwaa}{\usebox{\boxaa}}.
\end{equation}
The locations of the twist operators are
\begin{equation}
\begin{aligned}
&z_1 = -a-t+i\fr{\beta}{4}, \ \ \ \ \  &\bar{z}_1 =-a+ t - i\fr{\beta}{4}, \\
&z_2 = a-t+i\fr{\beta}{4}, \ \ \ \ \  &\bar{z}_2 = a+t - i\fr{\beta}{4}, \\
&z_3 = a-t-i\fr{\beta}{4}, \ \ \ \ \  &\bar{z}_3 = -a+t + i\fr{\beta}{4}, \\
&z_4 = a-t-i\fr{\beta}{4}, \ \ \ \ \  &\bar{z}_4 = a+t + i\fr{\beta}{4}. \\
\end{aligned}
\end{equation}
There are two candidates for the approximation of the correlation function.
We should choose the larger one from the two channels.
Each two-point function can be evaluated in the same way as the previous calculations with the following conformal maps;
\begin{enumerate}[(a)]

\item Mapping from cylinder to sphere: $z \to \ex{\fr{2\pi z }{\beta}}$.

\item Rotation: $z \to \fr{z-1}{z+1} $.

\item Mapping from sphere to cylinder $z \to \ln z$.

\end{enumerate}
Note that we need to care about the conformal factor unlike the previous calculations
because we have two limits, $\beta \to 0$ and $\epsilon \to 0$.
Considering going back to BCFT from the twist operator picture,
computing the conformal factor is equivalent to tracking how the size of the cutoff changes under the conformal transformations.
Taking the conformal factor into account, we obtain the following behavior up to constant.
\begin{equation}\label{eq:global1}
\begin{aligned}
S_A(t) &=\left\{
    \begin{array}{ll}
     \fr{4\pi c_{\text{eff}}}{3 \beta} t + \fr{2c_{\text{eff}}}{3} \log \fr{\beta}{2\pi \epsilon}  ,& \text{if }  0<t<a ,\\
     \fr{4\pi c_{\text{eff}}}{3 \beta} a + \fr{2c_{\text{eff}}}{3} \log \fr{\beta}{2\pi \epsilon}    ,& \text{if }  a<t .\\
    \end{array}
  \right.\\
\end{aligned}
\end{equation}

In comparing (\ref{eq:global1}) and  (\ref{eq:global2}),
we can see that the presence of an interface modifies the entanglement entropy of the system by replacing $c$ with $c_{\text{eff}}$. 
This can be interpreted as the fact that when quasiparticles hit the interface, only a portion of them can transmit through (as shown in the bottom of Figure \ref{fig:epr}), 
which implies that the entanglement entropy decreases in the presence of the interface \cite{Wen2018}.
In other words, $c_{\text{eff}}/c$ acts as a transmission coefficient for Bell pairs in the quasiparticle picture.

\subsection{Setup (ii'')}
In a similar manner, 
one can evaluate the entanglement entropy for the setup (ii'') in the high-temperature limit.
Alternatively, one can also evaluate the entanglement entropy by just the folding trick.
The result is as follows,
\begin{equation}
\begin{aligned}
S_A(t) =\left\{
    \begin{array}{ll}
     \fr{c_1+c_2}{6}  \fr{4\pi }{ \beta} t + \fr{c_1+c_2}{3} \log \fr{\beta}{2\pi \epsilon}  ,& \text{if }  0<t<a ,\\
   \fr{c_1+c_2}{6}    \fr{4\pi }{ \beta} a + \fr{c_1+c_2}{3} \log \fr{\beta}{2\pi \epsilon}    ,& \text{if }  a<t .\\
    \end{array}
  \right.\\
\end{aligned}
\end{equation}
This result looks similar to (\ref{eq:EEii}).
That is, the coefficient of the leading term is given by $(c_1+c_2)/6$ in the same way as (\ref{eq:EEii}).
This result can also be explained by the quasiparticle picture.

\subsection{Setup (iii'')}

\begin{figure}[t]
 \begin{center}
  \includegraphics[width=10.0cm,clip]{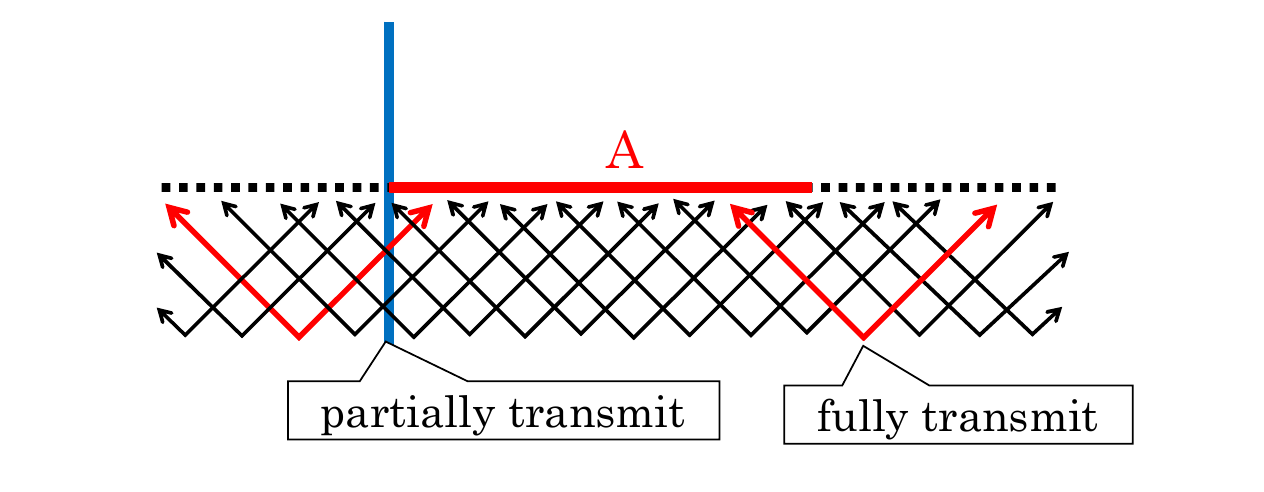}
 \end{center}
 \caption{Quasiparticule picture for the entanglement entropy in the system (iii'').
 The entanglement entropy counts two types of EPR pair:
 one is the EPR pair across the interface,
 and the other is the EPR pair in the CFT${}_1$.
 }
 \label{fig:epr2}
\end{figure}

In the high-temperature limit, the factorization still holds for the setup (iii'').
We can obtain the entanglement entropy as
\begin{equation}
\begin{aligned}
S_A(t) =\left\{
    \begin{array}{ll}
     \fr{c_{\text{eff}}+c_1}{6}  \fr{4\pi }{ \beta} t + \fr{c_{\text{eff}}+c_1}{3} \log \fr{\beta}{2\pi \epsilon}  ,& \text{if }  0<t<t_* ,\\
   \fr{c_1}{3}    \fr{4\pi }{ \beta} a + \fr{c_{\text{eff}}+c_1}{3} \log \fr{\beta}{2\pi \epsilon}    ,& \text{if }  t_*<t ,\\
    \end{array}
  \right.\\
\end{aligned}
\end{equation}
where $t_*=\fr{2c_1}{c_1+c_{\text{eff}}}$.
This is the generalized version of the formula (\ref{eq:conj}) that we expected.
We can again understand this result from the quasiparticle picture.
In the case (iii''), we have two types of EPR pairs:
one is the pair across the interface and the other is the pair in CFT${}_1$ (see Figure \ref{fig:epr2}).
This mixing leads to the coefficient $ \pa{c_{\text{eff}}+c_1}/6 $ of the entanglement entropy.
On the other hand, the entanglement entropy is saturated by the fully transmissive EPR pairs in the late time. Therefore, the coefficient is replaced by $c_1/3$.

\section{Universal Bound on Effective Central Charge}
\label{sec:bound}
Since we consider the most general interface and the effective central charge depends highly and non-trivially on the details of the interface,
we do not expect that the effective central charge has a universal form.
Nevertheless,
one may be able to give a {\it universal bound} on the effective central charge.
From the quasiparticle picture for the global quench,
it is natural to interpret the ratio $c_{\text{eff}}/\min(c_1,c_2)$ as the transmission coefficient.
Thus, we may expect
\begin{equation}
c_{\text{eff}} \leq \min(c_1,c_2).
\label{eq:boundagain}
\end{equation}
This bound seems to be reasonable because the partially transmissive interface should reduce entanglement between CFT${}_1$ and CFT${}_2$ and then decrease the entanglement entropy.
In this section, we will give a rigorous proof of this bound.

\subsection{Universal bound from entropic $c$-theorem}

\begin{figure}[t]
 \begin{center}
  \includegraphics[width=7cm,clip]{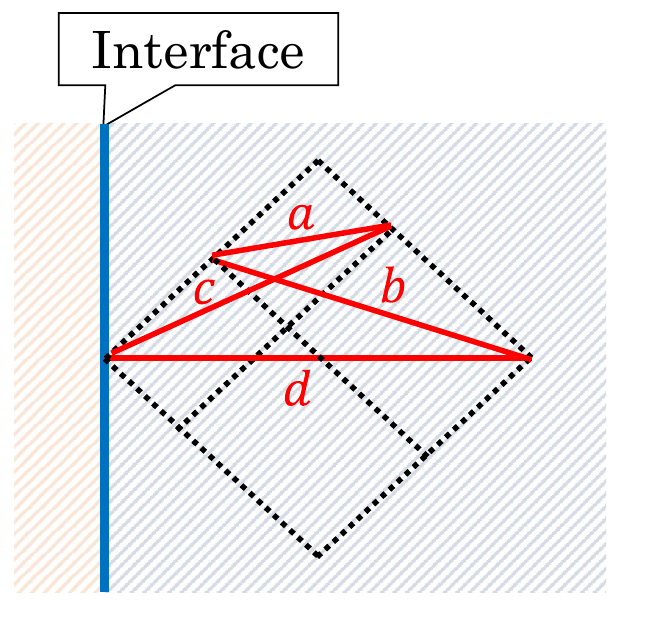}
 \end{center}
 \caption{
The vertical and horizontal axes correspond to the time direction and spatial direction, respectively.
The intersection of the domains of dependence of $b$ and $c$ is $a$, and their union is $d$.
The thick blue line represents the interface. The interface is in contact with one end of the interval $d$.
}
 \label{fig:causal}
\end{figure}

Here, we consider using an idea similar to the one used in the proof of the entropic $c$-theorem \cite{Casini2007}.
Let us consider two boosted intervals of length $\abs{b},\abs{c}$ respectively as sketched in Figure \ref{fig:causal}.
We label the intersection of the domains of dependence of $b$ and $c$ by an interval $a$ of length $\abs{a}$ and their union by an interval $d$ of length $\abs{d}$.
Using the strong subadditivity (SSA) for the entanglement entropy,
we have
\begin{equation}
\label{eq:SSA}
S(b)+S(c)\geq S(a)+S(d).
\end{equation}
By construction,
we have a simple relation,
\begin{equation}
\label{eq:relative}
\abs{a}\abs{d}=\abs{b}\abs{c}.
\end{equation}
The point is that (\ref{eq:SSA}) and (\ref{eq:relative}) still hold even when interfaces exist.
Consider placing an interface such that it touches one end of the interval $d$ (see Figure.\ref{fig:causal}).
The entanglement entropy for each interval can be evaluated using the results in Section \ref{sec:uni}.
In order to neglect the contributions from the constant terms,
we consider the limit $\abs{d}/\abs{a} \gg 1 $.
Using (\ref{eq:relative}) and the results about setup (iii) in Section \ref{sec:uni}, we obtain from (\ref{eq:SSA}),
\begin{equation}
\fr{c_1-c_{\text{eff}}}{6} \log \fr{\abs{b}}{\abs{a}} \geq 0.
\label{eq:ineq}
\end{equation}
Since $\abs{b}/\abs{a}>1$, the coefficients should be larger than or equal to zero, which implies
\begin{equation}
c_{\text{eff}} \leq c_1.
\end{equation}
Furthermore, by swapping CFT${}_1$ and CFT2${}_2$, we obtain
\begin{equation}
c_{\text{eff}} \leq c_2.
\end{equation}
Consequently, we obtain the universal bound on the effective central charge,
\begin{equation}
c_{\text{eff}} \leq \min(c_1,c_2).
\end{equation}
While there are various configurations (e.g., with two interfaces) where we obtain the same conclusion,
this one-interface configuration is optimal for a possible higher-dimensional case of Casini--Huerta, and it will be discussed in Section \ref{sec:disc}.

\subsection{Universal bound from holographic $c$-theorem}
\label{holo}

\begin{figure}[t]
 \begin{center}
  \includegraphics[width=8.0cm,clip]{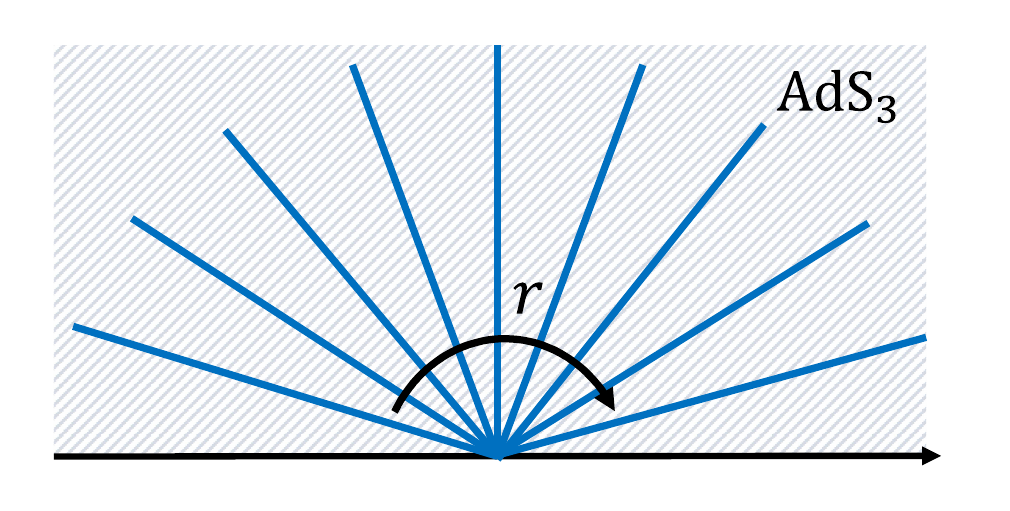}
 \end{center}
 \caption{The black line below corresponds to the AdS asymptotic boundary. Each blue line represents an AdS${}_2$ slice.}
 \label{fig:slice}
\end{figure}

In this subsection, we give an alternative proof of the universal bound using the AdS/ICFT correspondence \cite{Azeyanagi2008}.
The holographic dual of a generic interface can be described by the following metric (see Figure \ref{fig:slice}),
\begin{equation}
ds_{\text{AdS}_3}^2 = \ex{2A(r)} \fr{dx^2 - dt^2}{x^2} + dr^2.
\end{equation}
This metric is partitioning AdS${}_3$ into slices of AdS${}_2$.
The warpfactor $A(r)$ obeys the modified $c$-theorem \cite{Karch2001},
\begin{equation}\label{eq:modified}
A''(r) \leq -k \, \ex{-2A(r)},
\end{equation}
where $k$ is the sign of the constant curvature of the $2d$ slice.
Note that this inequality comes from the null energy condition.
By considering the case of $k=0$, that is flat spacetimes on each constant $r$ slice, this inequality can be applied to the proof of the standard holographic $c$-theorem \cite{Freedman:1999gp,Girardello:1998pd}.
In the AdS-sliced case, $k=-1$, the right-hand side is positive, making this a weaker condition compared to the flat case.
Nevertheless, this inequality is useful for our purpose as we will see below.

The relationship between the warpfactor and each central charge has been determined in \cite{Karch2021} as
\begin{equation}\label{eq:cs}
\begin{aligned}
c_1 &= \fr{3}{2G_N} \fr{1}{A'(\infty)}, \\
c_2 &= -\fr{3}{2G_N} \fr{1}{A'(-\infty)}, \\
c_{\text{eff}} &= \fr{3}{2G_N} \ex{A(r_{\text{min}})} , \\
\end{aligned}
\end{equation}
where $r_{\text{min}}$ is the $r$ that leads to the minimum warpfactor.
When approaching infinity, we reach AdS${}_3$ again, and the warpfactor reverts to its cosh form.
It is important to note that this implies that $\ex{-A}$ approaches zero at both positive and negative infinity.

We now have everything needed for the proof.
For convenience, we define
\begin{equation}\label{eq:F}
F(r) \equiv A'(r)^2 + \ex{-2A(r)},
\end{equation}
whose derivative is given by
\begin{equation}
F'(r) = 2\pa{A''(r) - \ex{-2A(r)} } A'(r),
\end{equation}
where the value inside the parentheses is always negative, which comes from the modified $c$-theorem (\ref{eq:modified}) with $k=-1$.
Using the facts for the warpfactor that we have demonstrated at the beginning of this subsection,
we can show the following properties of $F(r)$,
\begin{itemize}

\item
$F(r)$ reaches its maximum at $r_{\text{min}}$, where $A'(r)$ vanishes and thus $F(r_{\text{min}}) = \ex{-2A(r_{\text{min}})}$.

\item
At positive and negative infinity, the second term of (\ref{eq:F}) is suppressed, therefore, we have $F(\pm \infty) = A'(\pm \infty)^2$.

\end{itemize}
These two facts imply
\begin{equation}
\ex{-2A(r_{\text{min}})} \geq  A'(\pm \infty)^2.
\end{equation}
Finally, using the relation (\ref{eq:cs}), we obtain
\begin{equation}
c_{\text{eff}} \leq \min(c_1,c_2).
\end{equation}
That is, the modified $c$-theorem with $k=-1$ can be applied to understanding the effective central charge.


\section{Higher-dimensional Effective Central Charge}

\label{sec:higherholo}
Let us consider AdS$_{d+1}$/ICFT$_{d}$ for dimension $d>2$. In this section we will use the definition \eqref{eq:higherceff} to give the universal relation \eqref{eq:dresultii}, \eqref{eq:dresultiii} for general $d$-dimensional holographic ICFTs, and prove the universal bound \eqref{eq:bound} on $c_{\text{eff}}$ and $c_1,c_2$.

Consider the bulk as slicing the AdS$_{d+1}$ into AdS$_{d}$ leaves. As mentioned in Section \ref{sec:intro}, the subject of interest is the entanglement entropy of some codimensional-1 subregions in the $t=0$ slice, and we want to extract the ``universal" $\log\epsilon$ term (even $d$) or constant term (odd $d$) from it. In order to do so we have to choose our AdS$_{d+1}$ geometry carefully \cite{Casini:2011kv,Uhlemann:2023oea}. Namely, the ICFT lives on $\mathbb{R}\times S^{d-1}$, and the global AdS$_{d}$ metric reads 
\begin{equation}    
ds_{\text{AdS}_d}^2=d\rho^2-\cosh^2\rho dt^2+\sinh^2\rho ds_{S^{d-2}}^2,
\end{equation}
where $ds_{S^{d-2}}^2$ is the metric of a unit-size $S^{d-2}$. The subregion wraps around this sphere. The interface is located at $\rho=\infty$, and we introduce UV cutoff on $\rho$ as $\rho_\epsilon=\log (2/\epsilon)$. The total AdS$_{d+1}$ metric is then
\begin{equation}
ds_{\text{AdS}_{d+1}}^2=\ex{2A(r)}ds_{\text{AdS}_{d}}^2+dr^2.
\end{equation}
 The warpfactor solves the Einstein equation, and approaches $\ex{A(-\infty)}=L_1\cosh(r/L_1)$ for the CFT$_1$ side and $\ex{A(\infty)}=L_2\cosh(r/L_2)$ for the CFT$_2$ side.
 
\bigskip
The sketch of holographic Ryu-Takayanagi (RT) surfaces corresponding to case (i), (ii), and (iii) subsystems defined in Section \ref{sec:uni} was depicted, respectively, in Figure 9(a), 9(b) and 9(c) of \cite{Uhlemann:2023oea}. For case (i), the minimal RT surface is described by $r(\rho)$, and its surface area is given by
 \begin{equation} 
 \mathcal{A}=\Omega_{d-2}\int_0^{\log\frac{2}{\epsilon}} d\rho ~\ex{(d-2)A(r)}\sinh^{d-2}\rho\sqrt{\ex{2A(r)}+r'(\rho)^2},
 \end{equation}
where $\Omega_{d-2}=\frac{2\pi^{\frac{d-1}{2}}}{\Gamma(\frac{d-1}{2})}$ is the volume of a unit-size $S^{d-2}$. Taking $r(\rho)\equiv r_{\text{min}}$ solves the minimal-area condition, where $A(r)$ reaches its global minimum at $r_{\text{min}}$.\footnote{There could be multiple minima for the warpfactor coming from certain top-down construction (corresponding to different ways of splitting the interface Hilbert space \cite{Uhlemann:2023oea}), but for our purpose of defining the ``effective central charge'' we only choose the ``canonical splitting'' where $A(r_{\text{min}})$ is the true global minimum. The value of $r_{\text{min}}$ for $A$ to reach its true minimum can also be non-unique, but it does not affect the value of the entanglement entropy.} After integration we reach\footnote{Note that the factor ``2" in \ref{eq:case1ceff} is due to our convention of adding up contributions to the entanglement entropy from both sides, as explained in \cite{Karch2021}.}
\begin{equation}
\begin{split}
     S_{EE}^{(i)}&=2\frac{\mathcal{A}}{4G_N^{(d+1)}}=\frac{\Omega_{d-2}~\ex{(d-1)A(r_{\text{min}})}}{4G_N^{(d+1)}}\left(\frac{c_{d-2}}{\epsilon^{d-2}}+\dots+\frac{(-1)^{\frac{d}{2}-1}\Gamma(d-1)}{2^{d-3}~\Gamma\left(\frac{d}{2}\right)^2}\log\frac{2}{\epsilon}+\mathrm{const.}\right)\quad \mathrm{even}\   d\\
     S_{EE}^{(i)}&=\frac{\Omega_{d-2}~\ex{(d-1)A(r_{\text{min}})}}{4G_N^{(d+1)}}\left(\frac{c_{d-2}}{\epsilon^{d-2}}+\dots+\frac{(-1)^{\frac{d-1}{2}}2^{d-2}~\Gamma\left(\frac{d-1}{2}\right)^2}{\Gamma(d-1)}+O(\epsilon)\right)\quad \mathrm{odd}\   d.\\
\end{split}   
\label{eq:case1ceff}
\end{equation}

\bigskip
The case (ii) RT surface is described by $\rho(r)$. We will only attempt to recover the entanglement entropy of case (ii) for even $d$, with logarithm divergent terms. Near the limit $r\rightarrow \pm\infty$ we introduce the cutoff $r_\epsilon=L_i\log\frac{2}{\epsilon}$ near the two boundary CFTs $i=1,2$. The RT surface is anchored at $\lim_{u\rightarrow\infty}\rho(r)=\rho_0$ at the boundary. The surface area is 
\begin{equation}
    \mathcal{A}=\Omega_{d-2}\int dr~\ex{(d-2)A(r)}\sinh^{d-2}\rho~\sqrt{1+\ex{2A(r)}\rho'(r)^2}.
    \label{eq:a1}
\end{equation}

The EOM is
\begin{equation}
    (d-2)\ex{-2A(r)}\cosh\rho+(d-2)\rho'^2\cosh\rho- A'(r)\rho'\left[d+(d-1)\ex{2A(r)}\rho'^2\right]\sinh\rho-\rho''\sinh\rho=0.
\end{equation}

Now we expand $\rho$ near the boundary in terms of $\ex{-\frac{2|r|}{L_i}}$ as $\rho(r)=\rho_0+\rho_1~\ex{-\frac{2|r|}{L_i}}+\rho_2~\ex{-\frac{4|r|}{L_i}}+\dots.$
From the above EOM we know the expansion coefficients $\rho_i$ expressed as functions of $\rho_0$:
\begin{equation}   \rho_1=-2~\mathrm{coth}~\rho_0;~~\rho_2=-2~\mathrm{coth}~\rho_0~\mathrm{csch}^2~\rho_0;~~ \dots .
\end{equation}
Plugging them into \eqref{eq:a1}, we extract the constant term which we can in turn integrate to the cutoff $u_\epsilon$ to get the $\log\epsilon$. The \textit{Mathematica} result after integration is
\begin{equation}    
S_{EE}^{(ii)}=\frac{\Omega_{d-2}}{4G_N^{(d+1)}}\left(\dots+\frac{(-1)^{\frac{d}{2}-1}\Gamma(d-1)}{2^{d-2}~\Gamma\left(\frac{d}{2}\right)^2}(L_1^{d-1}+L_2^{d-1})\log\frac{2}{\epsilon}+\mathrm{const.}\right).
\label{eq:case2c}
\end{equation}

The odd $d$ situation for case (ii) entanglement entropy is more subtle since the constant term in principle depends on the details of the entire RT surface instead of just on the boundary sides close to the UV cutoff. This is different from the case (i), where we solved the entire RT surface to be flat, and were able to derive the universal constant in odd $d$\footnote{As explained in Section \ref{sec:intro} and below, we can still define $c_1,c_2$ in odd $d$ by calculating either $\langle TT\rangle$ or the holographic entanglement entropy of a narrow strip far away from the interface, deep inside CFT$_1$ and CFT$_2$ \cite{Myers:2010tj}. This is to be distinguished from the entanglement entropy of the interval at hand that intersects with the interface since the interface degrees of freedom affect the shape of the RT surface in case (ii) and (iii), and in turn the constant term.} Nevertheless, we can still proceed to find the bound for $c_{\text{eff}}$ defined from the case (i) in all dimensions $d>2$.


For the case (iii) interval, we would expect that one of its boundaries that is anchored on the interface contributes to the entanglement entropy the same as that of the case (i), similar to the 2d cases above. The reasoning is that assuming that the warpfactor $A(r)$ reaches its minimum at $r_{\text{min}}$, then from a standard near-boundary analysis, the generic RT surfaces satisfying the EOM will satisfy the same boundary condition $\lim_{\rho\rightarrow\infty}r(\rho)=r_{\text{min}}+O(e^{-(2+\delta)\rho})$ near the interface with some positive number $\delta>0$. The other contribution of the case (iii) depends on which CFT the subsystem is located, and for even $d$, it is either $c_1$ for the CFT$_1$ side, or $c_2$ for CFT$_2$, as the contribution to the logarithm divergent term only depends on the boundary of the interval. Hence we have the universal relation \eqref{eq:dresultiii} for even dimensional holographic ICFT.

\bigskip
For even $d$, we know that $L^{d-1}/4G_N^{(d+1)} \propto c$ in AdS$_{d+1}$/CFT$_{d}$ for the A-type trace anomaly $c$ \cite{Solodukhin:2008dh,Herzog:2015ioa}, which are precisely  $c_1$ in CFT$_1$ and $c_2$ in CFT$_2$. Explicitly, one can write down the Weyl anomaly for the $d$ dimensional CFT$_1$ as (same for CFT$_2$)
\begin{equation}
    \langle T_a^a\rangle=\sum_i B_iI_i+2(-1)^{\frac{d}{2}-1}c_1E_{(d)}+B'\nabla_aJ^a,
    \label{eq:weyl}
\end{equation}
where $E_{(d)}$ is the Euler density, and $I_i$ are independent Weyl invariants. 

Recall that near the boundary the AdS radius $L_1$ and $L_2$ can be expressed in terms of the warpfactor $A(r)$ as $1/L_1=A'(-\infty)$ and $1/L_2=A'(\infty)$. Following \cite{Myers:2010tj} and using the approach of \cite{Imbimbo:1999bj,Schwimmer:2008yh}, for even $d$ we have the expression for the A-type anomaly of CFT$_1$ and CFT$_2$ as:
\begin{equation}
    \begin{split}
        c_{1}&=\frac{\pi^{\frac{d}{2}}}{\Gamma\left(\frac{d}{2}\right)4\pi G_N^{(d+1)}(A'(-\infty))^{d-1}},\\
        c_{2}&=\frac{\pi^{\frac{d}{2}}}{\Gamma\left(\frac{d}{2}\right)4\pi G_N^{(d+1)}(A'(\infty))^{d-1}}.\\
    \end{split}
    \label{eq:ddc1c2}
\end{equation}

For odd $d$, we can define $c_1$ and $c_2$ by performing the standard procedure of analyzing the short-distance behavior of the stress tensor correlator $\langle TT\rangle$ far away from the interface. The resulting central charges are given by the same expressions \eqref{eq:ddc1c2} in terms of the warpfactor in the bulk \cite{Myers:2010tj}. 
 
By comparing the above equations to \eqref{eq:case2c}, using the $\Gamma$-function identity
\begin{equation}
    \Gamma(d-1)=\frac{1}{\sqrt{\pi}}~2^{d-2}~\Gamma\left(\frac{d-1}{2}\right)\Gamma\left(\frac{d}{2}\right),
    \label{eq:gammaid}
\end{equation}
we prove \eqref{eq:dresultii} for even $d$, which holds for any even-dimensional holographic ICFT, and is an interesting result on its own.

On the other hand, compare the definition of $c_{\text{eff}}$ in \eqref{eq:higherceff} and \eqref{eq:case1ceff}, and using \eqref{eq:gammaid} again, we have for both even and odd $d$,
\begin{equation}
\begin{split}
    c_{\text{eff}}&=\frac{\pi^{\frac{d}{2}}}{\Gamma\left(\frac{d}{2}\right)}\frac{\ex{(d-1)A(r_{\text{min}})}}{4\pi G_N^{(d+1)}}.\\
    \label{eq:ceffevenodd}
\end{split}   
\end{equation}
The rest are parallel arguments as in Section \ref{holo}. The inequality \eqref{eq:modified} for $A$ holds true for any dimensions. So in the same spirit, for either even or odd $d$, we have $$\ex{A(r_{\text{min}})}\le \mathrm{min}\left(A'(\infty)^{-1},A'(-\infty)^{-1}\right).$$ Comparing \eqref{eq:ceffevenodd} to \eqref{eq:ddc1c2}, we finally prove the bound \eqref{eq:bound} on $c_{\text{eff}}$ for holographic ICFT of any dimension greater or equal to $2$.

We leave a discussion on possible entropic proof of this bound using a generalized Casini--Huerta setup to the next section.

\section{Discussion}
\label{sec:disc}
We propose some remaining questions and directions for interesting future work in this section.
\begin{itemize}

\item \textit{Entropic proof of the bound on higher-dimensional} $c_{\text{eff}}$

It would be very interesting to formulate an entropic proof for the bound \eqref{eq:bound} for $c_{\text{eff}}$ in higher-dimensional ICFTs as defined in Section \ref{sec:higherholo}. The Casini--Huerta style approach used in Section \ref{sec:bound} is tempting, yet the single entangling surface (coinciding with the interface) bounding the hemisphere subregion (with a disc topology) in Section \ref{sec:higherholo} poses a challenge, which we sketch below.

We choose the standard metric on spacetime $\mathbb{R}_t\times S^{d-1}$, where the CFT lives, to be 
	\begin{equation}
		ds^2=-dt^2+d\phi_1^2+\sum_{a=2}^{n+1}\left(\prod_{m=1}^{a-1}\sin^2\phi_m\right)d\phi_a^2,\quad \phi_1\in[0,\pi],\quad\phi_2,\dots,\phi_{n+1}\in[0,2\pi].
	\end{equation}
The single interface $\mathbb{R}_t\times S^{d-2}$ is on the equator, namely at $\phi_1=\pi/2$. Now if we were to consider a spatial hyper-strip $A$ bounded by the equator and the ``Tropic of Cancer (or Capricorn)'' close to the equator, then we could prove an analogue of \eqref{eq:relative} for general $d$. Exploiting the rotational symmetry in the direction of $\phi_{d-1}$ by fixing it, we only need to focus on the $(d-2)$-dimensional cross-section (whose topology is $\mathbb{I}_{\phi_1}\times S^{d-3}$, parametrized by $\phi_1,\dots,\phi_{d-2}$) of the strip $A$, whose topology is $\mathbb{I}_{\phi_1}\times S^{d-2}$. Denote the width of $A$, in the direction of $\phi_1$, to be $|d|$. To mimic Figure \ref{fig:causal}, we boost $A$ in the direction of $\phi_1$ to get new hyper-strips with widths $|b|$ and $|c|$, and subsequently construct a hyper-strip with width $|a|$. All of them share the same unboosted $S^{d-3}$ in their cross-sections, so we only need to be concerned with the linear widths $|a|,\dots,|d|$. It turns out that in the most general case, we can prove the equality $|a||d|=|b||c|$.
 
However, one problem with this geometry is that its boundary other than the equator picks up the extrinsic curvature $K$ in the second term in Footnote \ref{foot:typec}, contributing to the C-type central charge, on top of the A-type one in \eqref{eq:weyl}. Hence, these hyper-strips are not directly helpful in the desired entropic proof of the bound on $c_{\text{eff}}$ defined in Section \ref{sec:higherholo}. We will leave the geometric details of hyper-strips and the new setup for the hemisphere subregion, as well as all subsequent entropic analysis to a future work.

We also note that the equality $\abs{a}\abs{d}=\abs{b}\abs{c}$ has also been generalized to arbitrary dimensions for \textit{strip} geometry on \textit{flat} CFT background in \cite{Myers:2012ed}.

\item \textit{More on higher-dimensional} $c_{\text{eff}}$

In this article, we proposed a higher-dimensional counterpart to the effective central charge from holographic viewpoint.
Given that it satisfies the same universal bound as in two dimensions,
it is expected to play a role as significant as the two-dimensional effective central charge.
Further analysis of this quantity, from both condensed matter theory and holography perspectives, would be an important future work.

\item \textit{Proof of universal bound from (\ref{eq:ceff})}

The effective central charge is defined in terms of the vacuum energy in the interface Hilbert space.
The modular bootstrap equation relates this vacuum energy to the CFT data at high energy.
For our purposes, the high-temperature picture would be useful, where the interface is a map between CFT${}_1$ and CFT${}_2$,
and the kernel of the map may play an important role in the decreasing of the entanglement. 
However, it is not straightforward to show the universal bound from (\ref{eq:ceff}).

One reason is that the bound on the vacuum energy and the bound on $c_{\text{eff}}$ are not directly related to each other.
Another obstruction comes from that a generic interface mixes different energy eigenstates.
For this reason, the partition function with interfaces does not have a simple form.

\item \textit{More universal properties of} $c_{\text{eff}}$

Until now, there has been no attempt to apply the ideas used in the proof of the $c$-theorem to studying ICFT.
However, our results have shown that this approach is extremely useful for investigating universal properties in ICFTs.

It would naturally be expected that similar techniques in different setups could lead to more findings of universal properties. Although straightforward, this will likely be an important direction for future work.

\item \textit{Other quantum information measures}

We have observed the universality of entanglement entropy in ICFT.
It is natural to expect similar universality in other quantum information quantities.
In particular, by examining the universality of measures of multipartite quantum correlations such as reflected entropy or negativity,
 we might obtain interesting implications for multipartite quantum correlations as in \cite{Kusuki2022b}. A very recent discussion of $c_{\text{eff}}$ using reflected entropy, based on both thin-brane holographic models as well as numerical studies, can be found in \cite{Tang:2023chv}.

\item \textit{Quenched states}

In this article, we investigated the universality of entanglement entropy for the vacuum state and the global quenched state in the presence of interfaces.
It would be interesting to see whether entanglement entropy also shows universality for other states such as the local quenched state.
Due to the presence of the interfaces, the entanglement is suppressed, and the behavior of information propagation becomes non-trivial.
This propagation behavior might allow for the classification of interfaces.
\end{itemize}

\section*{Acknowledgments}
We thank  Constantin Bachas, Horacio Casini, and  Gonzalo Torroba for useful discussions.
AK, HS, and MW are supported in part by the U.S. Department of Energy under Grant No. DE-SC0022021 and a grant from the Simons Foundation (Grant 651440, AK).
The work by YK and HO is supported in part by the U.S. Department of Energy, Office of Science, Office of High Energy Physics, under Award Number DE-SC0011632. In addition, YK is supported by the Brinson Prize Fellowship at Caltech.
HO is supported in part by the Simons Investigator Award (MP-SIP-00005259), the World Premier International Research Center Initiative, MEXT, Japan, and
JSPS Grants-in-Aid for Scientific Research 20K03965 and 23K03379. 
This work was performed in part at
the Aspen Center for Physics, which is supported by NSF grant PHY-1607611.

\clearpage
\bibliographystyle{JHEP}
\bibliography{main}

\end{document}